\documentclass[conference]{IEEEtran}

\usepackage{pifont}
\usepackage[utf8]{inputenc}
\usepackage{graphicx}
\usepackage{subfigure}
\usepackage{xcolor}
\usepackage{comment}
\usepackage{amsmath,amssymb,amsfonts}
\usepackage{booktabs}
\usepackage[justification=centering]{caption}
\usepackage{makecell}
\usepackage{indentfirst}
\usepackage{shorttoc}
\usepackage{multirow}
\usepackage{caption}

\definecolor{deepred}{rgb}{0.631,0.102,0.102}
     
\usepackage[colorlinks,linkcolor=deepred]{hyperref}

\hypersetup{
     colorlinks = true,
     urlcolor = deepred
     }

\usepackage[linesnumbered, ruled]{algorithm2e} 
\usepackage{algorithmic}

\SetCommentSty{mycommfont}

\newenvironment{packeditemize}{
\begin{list}{$\bullet$}{
\setlength{\labelwidth}{8pt}
\setlength{\itemsep}{0pt}
\setlength{\leftmargin}{\labelwidth}
\addtolength{\leftmargin}{\labelsep}
\setlength{\parindent}{0pt}
\setlength{\listparindent}{\parindent}
\setlength{\parsep}{0pt}
\setlength{\topsep}{3pt}}}{\end{list}}

\newcommand{\REQ}[2]{\vspace{3pt}\noindent\fbox{\parbox{0.97\linewidth}{\noindent{\textbf{{Property \##1:}}} #2}}\vspace{3pt}}
\newcommand{\qiuhan}[1]{\textbf{\textcolor{blue}{[Han: #1]}}}
\newcommand{\tianwei}[1]{\textbf{\textcolor{red}{[Tianwei: #1]}}}

\newcommand{\yi}[1]{\textbf{\textcolor{deepred}{[Yi: #1]}}}
\IEEEoverridecommandlockouts
\begin{document}

\title{Mitigating Advanced Adversarial Attacks with More Advanced Gradient Obfuscation Techniques}


\author{
\IEEEauthorblockN{
Han Qiu\IEEEauthorrefmark{1}$^{1}$,
Yi Zeng\IEEEauthorrefmark{2}$^{1}$,
Qinkai Zheng\IEEEauthorrefmark{3}, 
Tianwei Zhang\IEEEauthorrefmark{4}, 
Meikang Qiu\IEEEauthorrefmark{5},
Gerard Memmi\IEEEauthorrefmark{1}
}
\IEEEauthorblockA{
\IEEEauthorrefmark{1}Telecom Paris, Institut Polytechnique de Paris, Palaiseau, France, 91120.\\ \IEEEauthorrefmark{2}University of California San Diego, CA, USA, 92122.\\ 
\IEEEauthorrefmark{3}Shanghai Jiao Tong University, Shanghai, China, 200240.\\ 
\IEEEauthorrefmark{4}Nanyang Technological University, Singapore, 639798.\\ 
\IEEEauthorrefmark{5}Columbia University, New York, USA, 10027.\\
\IEEEauthorrefmark{1}\{han.qiu, gerard.memmi\}@telecom-paris.fr,
\IEEEauthorrefmark{2}y4zeng@eng.ucsd.edu, 
\IEEEauthorrefmark{3}paristech-hill@sjtu.edu.cn,\\
\IEEEauthorrefmark{4}tianwei.zhang@ntu.edu.sg,
\IEEEauthorrefmark{5}qiumeikang@yahoo.com}

}

\maketitle

\begin{abstract}


\emph{Deep Neural Networks}~(DNNs) are well-known to be vulnerable to \emph{Adversarial Examples}~(AEs). A large amount of efforts have been spent to launch and heat the arms race between the attackers and defenders. Recently, advanced gradient-based attack techniques were proposed (e.g., BPDA and EOT), which have defeated a considerable number of existing defense methods. Up to today, there are still no satisfactory solutions that can effectively and efficiently defend against those attacks.

In this paper, we make a steady step towards mitigating those advanced gradient-based attacks with two major contributions. First, we perform an in-depth analysis about the root causes of those attacks, and propose four properties that can break the fundamental assumptions of those attacks. Second, we identify a set of operations that can meet those properties. By integrating these operations, we design two preprocessing functions that can invalidate these powerful attacks. Extensive evaluations indicate that our solutions can effectively mitigate all existing standard and advanced attack techniques, and beat 11 state-of-the-art defense solutions published in top-tier conferences over the past 2 years. The defender can employ our solutions to constrain the attack success rate below 7\% for the strongest attacks even the adversary has spent dozens of GPU hours.

\end{abstract}

\footnotetext[1]{equal contribution}

\section{Introduction}
In the year of 2013, Szegedy et al. \cite{szegedy2013intriguing} proposed the concept of \emph{Adversarial Examples}~(AEs) to attack \emph{Deep Learning}~(DL) models: with imperceptible and human unnoticeable modifications to the input, the model will be fooled to give wrong prediction results. 
This study has received widespread attention from both machine learning and security communities. 
Such popularity is due to two reasons. 
First, the existence of AEs reveals the distinctions of understanding and interpretation between humans and machines. 
Although DL models have satisfactory performance in terms of automation, speed, and possibly accuracy, they can make simple mistakes that humans will never do. 
So understanding the mechanisms of AEs can help improve the robustness of DL models. 
Second, as DL techniques have been widely commercialized in various products, AEs can threaten these DL applications and bring catastrophic consequences to our daily life. 
The practicality and severity of AEs have been proved by past works (autonomous driving \cite{cao2019adversarial,zhao2019seeing,eykholt2018robust}, home automation \cite{carlini2016hidden,yuan2018commandersong,zhang2017dolphinattack}, etc.). 
It is in urgent need of effective and efficient defense solutions to thwart AEs. 

\begin{figure}[t]
  \centering
  \includegraphics[width=\linewidth]{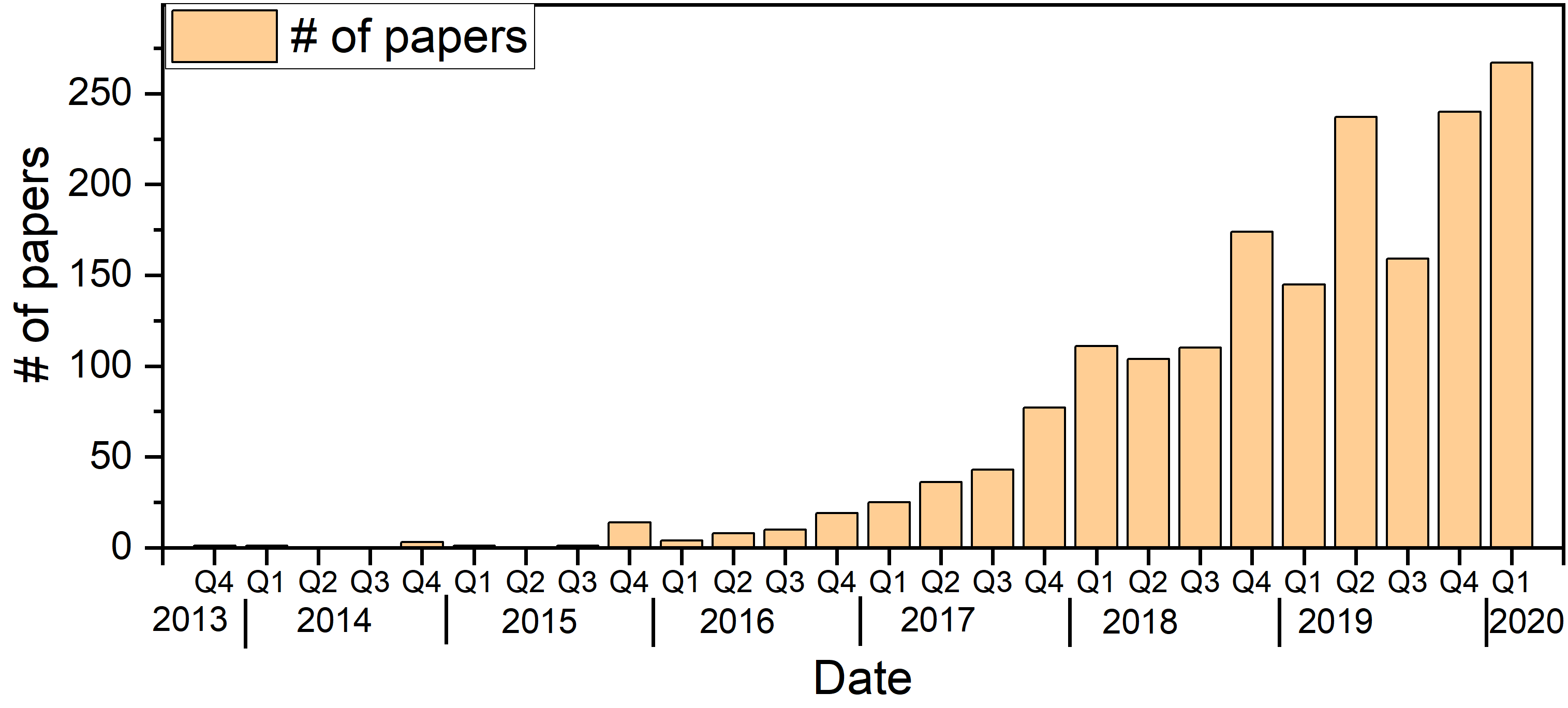}
  \caption{The number of research papers published on Arxiv.org about adversarial examples. Data source: \cite{papernumber}}
  \label{fig:paper-count}
\end{figure}

Over the past seven years, a great number of research papers have been published in this topic, and it is still growing at an increasing speed (\figurename~\ref{fig:paper-count}). 
A majority of these papers focused on enhancing the powers of AEs or mitigating the new attack techniques. 
This leads to an arms race between attacks and defenses. 
Generally speaking, the generation of AEs can be converted into an optimization problem: searching for the minimal perturbations that can cause the model to predict a different targeted or untargeted label. 
Then, common attack techniques adopt the gradient-based approaches to identify the optimal perturbations (e.g., FGSM~\cite{goodfellow2014explaining}, I-FGSM~\cite{kurakin2016adversarial}, LBFGS~\cite{szegedy2013intriguing}, DeepFool~\cite{moosavi2016deepfool}, and C\&W~\cite{carlini2017towards}).
To defeat those attacks, a lot of defense solutions were proposed to obfuscate the gradients and increase the difficulty of solving the optimization \cite{guo2018countering,prakash2018deflecting,xie2018mitigating,buckman2018thermometer}. 
These solutions tried to either prohibit the gradient calculation by introducing non-differentiable operations (shattered gradients) or randomize the gradient to hide the actual value (stochastic gradients).

Unfortunately, neither of the two directions can provide strong protection and robustness, and advanced attacks were proposed to break those gradient obfuscation based defenses 
\cite{athalye2018obfuscated,athalye2018synthesizing}. 
Specifically, to cope with shattered gradients, \emph{Backward Pass Differentiable Approximation}~(BPDA) was introduced to approximate the gradient values by replacing the non-differentiable function with its input on the backward pass. 
To handle stochastic gradients, \emph{Expectation over Transformation}~(EOT) was designed to calculate the gradient of random functions as the average values of multiple sessions statistically. 
These advanced techniques have successfully defeated the previous defensive solutions \cite{athalye2018obfuscated}. 
Even after their disclosure, new defense methodologies published in the recent two years are still proven to be vulnerable to either BPDA, EOT, or their combination \cite{tramer2020adaptive}. 
To the best of our knowledge, there are still no satisfactory preprocessing-based approaches to mitigate those two advanced techniques up to today. They seem to have ended the arms race with the adversary's victory.

The question we want to address in this paper is: \emph{is it possible to continue the competition with more advanced defense solutions to mitigate the above advanced attacks?} 
This is an extremely challenging task due to the omnipotence of those attacks. On one hand, they assume very high adversarial capabilities: the attacker knows every detail (mechanisms and parameters) of the DL model as well as the potential defenses. This assumption can significantly increase the difficulty for defense designs, and existing solutions that require to hide the model or defense mechanisms are not applicable. 
On the other hand, BPDA and EOT techniques target the root causes of gradient obfuscation: the introduced mitigation operations need to be carefully designed and configured in order not to affect the model accuracy. As such, the non-differentiable operation can always be approximated as the input, and the random operation can always be estimated by averaging over randomization. Based on these two observations, the attacks can effectively break existing gradient obfuscation techniques.

We propose defense solutions to overcome the above challenges and increase the difficulty and cost of these advanced adversarial attacks. The key insight of our solutions is that the assumption in BPDA (the preprocessing function $g(\cdot)$ must satisfy the property $g(x)\approx x$ in order to guarantee $f(g(x))=f(x)$, where $f$ is the target model) is not necessary and can be relaxed.

Our first contribution is to propose four properties for designing a preprocessing function $g(\cdot)$ to mitigate those attacks: (1) $g(\cdot)$ can maintain the model accuracy: $f(g(x))=f(x)$; (2) it has relatively big divergence from $x$; (3) its gradient is random and unpredictable; (4) it is non-differentiable. The first three requirements can invalidate BPDA attacks, while the last one can invalidate EOT. It can definitely mitigate other standard adversarial attacks as well.

Our second contribution is to identify a set of operations that meet these properties: Feature Distillation (\texttt{FD}) \cite{liu2019feature} is an advanced compression-based method. Randomization Layer (\texttt{Rand}) \cite{xie2018mitigating}
preprocesses the image with random rescaling, padding, and reshaping. We further design a new operation, Random Distortion over Grids (\texttt{RDG}), which distorts the image via random pixel dropping and displacement. We integrate these operations into two preprocessing functions (\texttt{FD}+\texttt{Rand} and \texttt{FD}+\texttt{RDG}) that can satisfy all the four requirements, which forms our defense solutions. 



We conducted extensive experiments to show the effectiveness of our solutions. Our solutions can constrain the attack success rate under 7\% even against 10000 rounds of BPDA+EOT attack, which significantly outperform 11 state-of-the-art gradient obfuscation defenses from top-tier conferences.

We expect that our solutions can heat the arms race of adversarial attacks and defenses, and contribute to the defender side. We hope the four properties can inspire researchers to come up with more robust defense strategies. Meanwhile, we also hope that adaptive attacks against our defenses can be designed. All these efforts can advance the study and understanding of adversarial examples and DL model robustness. To better promote this research direction, we release a toolkit online\footnote{\url{https://github.com/YiZeng623/Advanced-Gradient-Obfuscating}}, including the implementation of our defense techniques, a summary of other defense methods as well as various adversarial attacks.

\section{Backgrounds}\label{backgrounds}

\subsection{Attack Concept and Scenarios}

An adversary can add human-unnoticeable perturbations on the original input to fool a DNN classifier. 
Formally, the target DNN model is a mapping function $f(\cdot)$. Given a clean input sample $x$, the corresponding AE is denoted as $\widetilde{x}=x+\delta$ where $\delta$ is the adversarial perturbation. 
Then AE generation can be formulated as the optimization problem in Equation \ref{eq:AE}a (targeted attack where $l'\neq f(x)$ is the desired label set by the attacker) or Equation \ref{eq:AE}b (untargeted attack). 

\begin{subequations}
\label{eq:AE}
  \begin{align}
& min  \lVert\delta\rVert, s.t. \: f(\widetilde{x})=l' \\
& min  \lVert\delta\rVert, s.t. \: f(\widetilde{x})\neq f(x)
  \end{align}
\end{subequations}

Generally, there are two attack scenarios~\cite{carlini2019evaluating}, determined by the adversary's knowledge about the target system.
(1) \emph{White-box scenario}: the adversary knows every detail about the neural network model including the architecture and all the parameters. He is also aware of the defense mechanism and the corresponding parameters. (2) \emph{Black-box scenario}: the adversary does not have any knowledge about the victim system. In addition to these two scenarios, there are also some works~\cite{prakash2018deflecting} assuming the adversary knows all details about the model but not the defense mechanism. It is not quite realistic and reasonable to hold the defense secret, as ``this widely held principle is known in the field of security as Kerckhoffs' principle.''~\cite{carlini2019evaluating}. So we exclude this scenario in this paper.

\subsection{Development History}
\label{sec:devep-hist}

\noindent{\textbf{Round 1: attack.}}
As the first study, Szefedt et al.~\cite{szegedy2013intriguing} adopted the L-BFGS algorithm to solve the optimization problem of AE generation. 
Shortly after this work, a couple of gradient-based methods were introduced to enhance the attack techniques: the gradient descent evasion attack~\cite{biggio2013evasion} calculated the gradients of neural networks to generate AEs; \emph{Fast Gradient Sign Method}~(FGSM)~\cite{goodfellow2014explaining} calculated the adversarial perturbation based on the sign of gradients, which was further improved by its iterative versions (I-FGSM~\cite{kurakin2016adversarial} and MI-FGSM~\cite{dong2017discovering}). 
Deepfool~\cite{moosavi2016deepfool} is another iterative method that outperforms previous attacks by searching for the optimal perturbation across the decision boundary. 
Meanwhile, some other techniques were proposed to increase the attack efficiency: \emph{Jacobian-based Saliency Map Attack}~(JSMA)~\cite{papernot2016limitations} estimated the saliency map of pixels w.r.t the classification output, and only modified the most salient pixels. 
One pixel attack~\cite{su2019one} is an extreme-case attack where only one pixel can be modified to fool the classifier. 

\noindent{\textbf{Round 2: defense.}}
With the advance of adversarial attacks, defense solutions were proposed to increase the robustness of DNN models. 
They can be classified into three categories. 
The first direction is adversarial training~\cite{kurakin2016adversarial, huang2015learning, shaham2018understanding}, where AEs are used with normal examples together to train DNN models to recognize and correct malicious samples. 
The second direction is to train other models to assist the target one. 
Magnet~\cite{meng2017magnet} used detector networks to identify AEs by approximating the manifold of normal examples. 
Generative Adversarial Trainer~\cite{lee2017generative} utilized training target networks along with a generative network to generate adversarial perturbation for the target model to distinguish. 
The third direction is to design AE-aware network architecture or loss function. Deep Contractive Networks~\cite{gu2014towards} added a contractive penalty to alleviate the effects of AEs. 
Input Gradient Regularization~\cite{ross2018improving} countered AEs by penalizing the degree of variations of input perturbations on the output. 
Defensive distillation~\cite{papernot2016distillation} generated soft training labels from one network and retrained a second network with higher robustness. 
This method claimed to have very high resistance against AEs and was one of the strongest defenses at that time.

\noindent{\textbf{Round 3: attack.}}
A more powerful attack, C\&W~\cite{carlini2017towards}, was proposed by updating the objective function to minimize $l_p$ distance between AEs and normal examples. 
C\&W can effectively defeat Defensive Distillation~\cite{carlini2017towards} and other defenses with assisted models~\cite{carlini2017magnet} with very high attack success rates. 


\noindent{\textbf{Round 4: defense.}}
Since then, new defense strategies were introduced to increase the difficulty of AE generations by obfuscating the gradients. 
Five input transformations were tested to counter AEs in~\cite{guo2018countering}, including image cropping and rescaling, bit-depth reduction, JPEG compression, total variance minimization~(TV), and image quilting. 
Prakash et al. \cite{prakash2018deflecting} designed \emph{Pixel Deflection}~(PD), which randomly redistributes a small number of pixels as artificial perturbation and applies wavelet-based denoising to remove both artificial and adversarial perturbation. 
Xie et al. \cite{xie2018mitigating} proposed to use a randomization layer to randomly rescale the input image with zero-paddings. 
Buckman et al. \cite{buckman2018thermometer} introduced Thermometer encoding, which encodes input images with discrete values to prevent the direct calculation of gradient descent during AE generation. 
Das et al. \cite{das2018shield} proposed SHIELD, a defense that compresses different regions of an image with random compression levels to mitigate AE perturbations.
Those solutions are effective against all prior attacks. 

\noindent{\textbf{Round 5: attack.}}
To particularly target the gradient obfuscation-based defenses, two more advanced attacks were introduced. 
BPDA~\cite{athalye2018obfuscated} copes with the non-differentiable obfuscation operation by approximating the gradients during back-propagation. 
EOT~\cite{athalye2017synthesizing} deals with the randomization obfuscation operation by averaging the gradients of multiple sessions. 
More detailed descriptions about BPDA and EOT can be found in Section \ref{sec:overview}. 
After the disclosure of these two attacks, a large number of defense works have been published. 
Unfortunately, most of them did not consider or incorrectly evaluate these two attacks, and some representative solutions have been analyzed and proved to be incapable of defeating BPDA and EOT attacks~\cite{tramer2020adaptive}. 
Up to now, there are still no effective preprocessing-based defenses. 
This is what we aim to address in this paper.

\section{Problem definition and Threat Model}

It is necessary to specify the adversarial capabilities and defense requirements in our consideration, as described below. 

\subsection{Threat Model}
\label{sec:threat-model}

\noindent{\textbf{Adversarial Goals.}}
There are two main types of adversarial attacks: untargeted attacks that try to mislead the DNN models to an arbitrary label different from the correct one, and targeted attacks which succeed only when the DNN model predicts the input as one specific label desired by the adversary \cite{carlini2017towards}. In this paper, we only evaluate the targeted attacks. The untargeted attacks can be mitigated in the same way. 

\noindent\textbf{{Adversary's Knowledge.}}
We consider a white-box scenario, where the adversary has full knowledge of the DNN model, including the network architecture, exact values of parameters, and hyper-parameters. We further assume that the adversary has full knowledge of the proposed defense, including the algorithms and parameters. For the defenses employing randomization techniques, we assume the random numbers generated in real-time are perfect with a large entropy such that the adversary cannot obtain or guess the correct values.

It is worth noting that this white-box scenario represents the strongest adversaries. 
Under such a scenario, a big number of existing state-of-the-art defenses are invalidated as shown in~\cite{tramer2020adaptive}. 
This also significantly increases the difficulty of defense designs. 

\noindent{\textbf{Adversarial Capabilities.}}
The adversary is outside of the DNN classification system, and he is not able to compromise the inference computation or the DNN model parameters (e.g., via fault injection to cause bit-flips \cite{rakin2019bit} or backdoor attacks \cite{gao2019strip}). All he can do is to manipulate the input data with imperceptible perturbations. In the context of computer vision tasks, he can directly modify the input image pixel values within a certain range. We use $l_{\infty}$ and $l_{2}$ distortion metrics to measure the scale of added perturbations: we only allow the generated AEs to have either a maximum $l_{\infty}$ distance of 8/255 or a maximum $l_{2}$ distance of 0.05 as proposed in~\cite{athalye2018obfuscated}. 

\subsection{Defense Requirements}
\label{sec:def-req}


Based on the above threat model, we list a couple of requirements for a good defense solution: 

First, there should be no modifications to the original DNN model, e.g., retraining a model with different structures~\cite{papernot2016distillation} or datasets~\cite{tramer2017ensemble}. We set this requirement for two reasons. (1) Model retraining can significantly increase the computation cost, especially for large-scale DNN models (e.g. ImageNet scale~\cite{imagenet_cvpr09}). (2) Those defense methods lack generality to cover various types of attacks. They ``explicitly set out to be robust against one specific threat model'' \cite{carlini2019evaluating}.

Second, we consider adding a preprocessing function over the input samples before feeding them into the DNN models. Such preprocessing operation can either remove the effects of adversarial perturbations on the inference or make it infeasible for the adversary to generate AEs adaptively, even he knows every detail of the operation. This function should be general-purpose and applicable to various types of data and DNN models of similar tasks. 

Third, this preprocessing function should be lightweight with negligible computation cost to the inference pipeline. Besides, it should also preserve the usability of the original model without decreasing its prediction accuracy. Input preprocessing can introduce a trade-off between security and usability: the side effect of correcting the adversarial examples can also alter the prediction results of clean samples. A qualified operation should balance this trade-off with maximum impact on the adversarial samples and minimal impact on the clean ones.

\section{Methodology Insights and Overview}
\label{sec:overview}

Our proposed solution is based on more advanced gradient obfuscation techniques. We propose to use a preprocessing function $g(\cdot)$, which can defeat the advanced attacks within the threat model in Section \ref{sec:threat-model} and satisfy the requirements in Section \ref{sec:def-req}. In this section, we analyze these attacks and identify the design philosophy of our methodology. We give specific examples of qualified functions in the next section. 

\subsection{Mitigation of Gradient Approximation Attacks}
\label{sec:method-bpda}

\noindent{\textbf{Attack analysis.}} As mentioned in Section~\ref{sec:devep-hist}, for shattered gradient based defense solutions, an non-differentiable operation $g(\cdot)$ is integrated with the target DNN model $f(\cdot)$ to preprocess the input samples. Then the model inference process becomes $y=f(g(x))$. 
This can increase the difficulty of calculating gradients directly since $g(\cdot)$ is non-differentiable. Thus AE generation becomes infeasible. 


To defeat such defense, BPDA attack was proposed~\cite{athalye2018obfuscated}. with a key assumption that the preprocessing function $g(\cdot)$ maintains the property $g(x)\approx x$ in order to preserve the functionality of the target model $f(\cdot)$. 
This assumption is held for all existing shattered gradient based solutions. Then the adversary can use $g(\cdot)$ on the forward pass and replace it with $x$ on the backward pass when calculating the gradients. The derivative of the preprocessing function will be approximated as the derivative of the identity function, which is 1. The gradient calculation process is described in Eq. \ref{eq:bpda1}. 

\begin{subequations}
\label{eq:bpda1}
  \begin{align}
\bigtriangledown _{x}f(g(x))| _{x = \hat{x}} & = \bigtriangledown _{x} f(x)| _{x = g(\hat{x})} \times \bigtriangledown _{x} g(x) | _{x = \hat{x}} \\
& \approx \bigtriangledown _{x} f(x)| _{x = g(\hat{x})} \times \bigtriangledown _{x} x | _{x = \hat{x}} \\
& = \bigtriangledown _{x} f(x)| _{x = g(\hat{x})}
  \end{align}
\end{subequations}

\noindent{\textbf{Attack mitigation.}}
We identify several necessary properties 
of the preprocessing function $g(\cdot)$ in order to mitigate the BPDA attack. First, this function must preserve the functionality of the target model, without affecting the prediction results of normal samples. This gives the first property for our solution:

\REQ{1}{$g(\cdot)$ cannot affect the prediction results: $f(g(x)) = f(x)$}.

The BPDA attack depends on the condition that $g(x) \approx x$. To invalidate such an attack, we need to design a preprocessing function that can break this assumption. This leads to the second property as below. Note that properties \#1 and \#2 are not contradictory. It is possible (although challenging) to identify such preprocessing functions that can satisfy both properties. We will show detailed examples in Section \ref{subsec:analysis}.

\REQ{2}{$g(x)$ has a large variance from $x$: $g(x) \not\approx x$}

We should also consider more adaptive attacks. 
Even the adversary cannot use $x$ to replace $g(x)$ for gradient calculation with properties \#1 and \#2, it is still possible that he might discover an alternative way to approximate $\bigtriangledown_{x}g(x)$ when $x = \hat{x}$. 
To eliminate such possibility, we should design a randomization function $g(\cdot)$ which is the third property: 


\REQ{3}{$g(\cdot)$ is a randomization function. A fixed input $\hat{x}$ leads to different outputs each time: $g(\hat{x}) \not\approx  const$.}

\subsection{Mitigation of Gradient Expectation Attacks}
 
\noindent{\textbf{Attack analysis.}} 
Another popular preprocessing operation is to randomize the gradient of the model in real-time without affecting the prediction results. To evade such defense strategy, EOT was adopted as an attack strategy~\cite{athalye2018obfuscated}, which statistically computes the gradients over the expected transformation to the input $x$. Formally, for a preprocessing function $g(\cdot)$ that randomly transforms $x$ from a distribution of transformations $T$, EOT optimizes the expectation over the transformation with respect to the input by $\mathbb{E} _{t\sim T}f(g(x))$, as shown in Eq. \ref{eq:eot1}. The adversary can always get a proper expectation with samples at each gradient descent step. 

\begin{equation}\label{eq:eot1}
    \bigtriangledown_{x}  \mathbb{E} _{t\sim T}f(g(x)) = \mathbb{E} _{t\sim T} \bigtriangledown_{x} f(g(x))
\end{equation}

\noindent{\textbf{Attack mitigation.}}
EOT needs to calculate the gradient of $f(g(x))$ in Eq. \ref{eq:eot1} to calculate the expectation over the transformation. It needs the requirement that $g(\cdot)$ is differentiable. This gives us the last property for our preprocessing strategy to mitigate EOT attack: 

\REQ{4}{$g(\cdot)$ is non-differentiable.} 

\subsection{Put Everything Together}
\label{sec:method-bpda+eot}

In reality, it is difficult to discover a single operation that can meet all four properties. To simplify this problem, we can concatenate two operations: $g = g_1 \circ g_2$. Operation $g_1(\cdot)$ focuses on the gradient approximation attack, with properties \#2 and \#3. Operation $g_2(\cdot)$ focuses on the gradient expectation attack, with property \#4. Both two operations have property \#1. The integration of them can mitigate advanced attacks (BPDA, EOT, or their combination), as well as classic attacks (e.g., FGSM, C\&W, Deepfool). We present the analysis below. 
 
\noindent{\textbf{Mitigating Classic Attacks.}} 
Operation $g_1(\cdot)$ can lead to stochastic gradients caused by the randomization factor. Operation $g_2(\cdot)$ can give shattered gradients. As a result, Function $g$ enjoys both of the two obfuscation features and can defeat classic attacks as other preprocessing-based solutions. 

\noindent{\textbf{Mitigating Advanced Attacks.}}
Due to the non-differentiable $g_2(\cdot)$, it is impossible to employ EOT only to compromise the function $g(\cdot)$. The adversary can ignore the existence of $g_2(\cdot)$ and use EOT to generate perturbations only from $f(\cdot)$ and $g_1(\cdot)$. Such AEs cannot fool the model expected by the adversary, as the operation $g_2(\cdot)$ can invalidate the perturbation at the inference stage. The adversary has to use a brute force way to break the $g_2(\cdot)$, which has an extremely high cost. Empirical evaluations will be demonstrated in Section \ref{sec:eval}.

Due to properties \#2 and \#3 of $g_1(\cdot)$, the BPDA technique cannot get the correct gradients and optimal perturbation. The integration of BPDA and EOT techniques can remove the obstruction from property \#3. However, property \#2 can still hinder the adversary from generating effective AEs. We can see this is the essential property to mitigate advanced attacks, which existing solutions do not have.
\section{Examples of Defense Operations}
\label{sec:technique}


In this section, we identify some candidate operations that can satisfy the properties proposed in Section \ref{sec:overview}.

\subsection{Operation $g_1(\cdot)$}
This needs to satisfy the properties of \#1, \#2, and \#3. We discover one qualified operation from existing work and also propose a new transformation, which is more efficient. 

\subsubsection{Randomization Layer (Rand)}
Xie et al. \cite{xie2018mitigating} introduced a randomization layer before the model to preprocess the image input. Without loss of generality, we consider an image size of $299 \times 299$ (ImageNet). Then the first step is to select a random number $r$ within the range [299, 330], and rescale the input image to the size of $r \times r$. Then it adopts a zero-padding technique to randomly add zeros to the four sides of the image and pad it to the size of $400 \times 400$. Before this input is fed to the target model (e.g., inception v3), it will be reshaped to $299 \times 299$ again. With such rescaling-padding-reshaping steps, certain information or pixels (including adversarial perturbation) will be dropped with a unified probability. 

Based on the evaluations in \cite{xie2018mitigating}, the introduction of such a randomization layer has negligible performance degradation to the target model (property \#1). Due to the large randomness in rescaling and padding, the operation causes a great variance between the original and preprocessed samples (property \#2). We will quantitatively validate this argument in Section \ref{subsec:analysis} and Table \ref{tab: compare}. Its gradients cannot be approximated as a constant (property \#3). Thus this is a qualified candidate for $g_1$.


\subsubsection{Random Distortion over Grids (\texttt{RDG})}

\begin{figure*}[h]
  \centering
  \includegraphics[width=\textwidth]{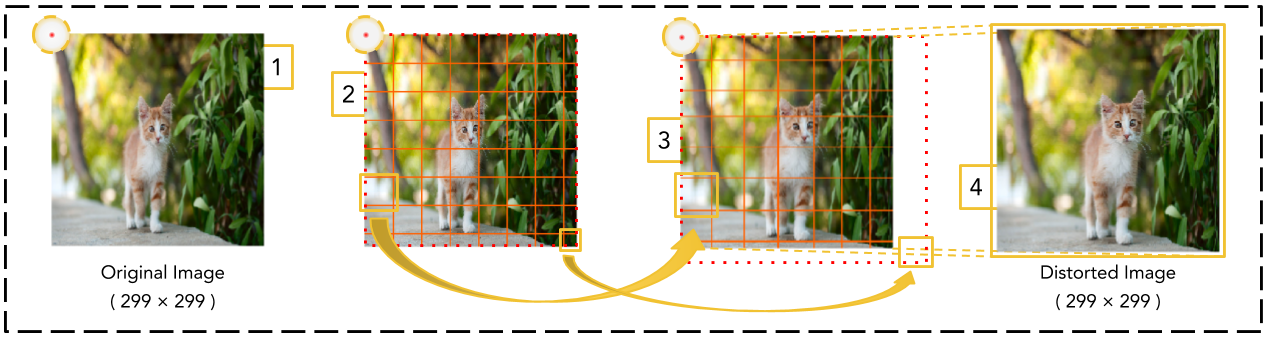}
  \caption{Four stages of the Random Distortion over Grids: (1) Randomly select a corner as the starting point; (2) Randomly generate $n$ grids vertically and horizontally from the starting point following the variation limits, $\delta$; (3) Distort between each neighboring grids, compress or stretch, to remap each pixel away from the original coordinates; (4) Acquire the distorted image of the same shape as the input.}
  \label{fig:RDG}
\end{figure*}

We propose a novel transformation method by adopting the image distortion method as the preprocessing function $g_1(\cdot)$. 
It can provide a greater variance between the original and transformed samples without affecting the model performance. 



Our solution originates from the dropping-pixels strategy \cite{xie2018mitigating,guo2018countering}. The general idea is to drop certain randomly selected pixels from the original image, and displace each pixel away from the original coordinates. The whole procedure of \texttt{RDG} consists of four stages, as illustrated in Fig.\ref{fig:RDG} and in Algorithm \ref{algo:RDG}. (1) One of the four corners is randomly selected as a starting point, e.g. the upper-left corner (line 1). (2) The original image is a randomly distorted grid by grid. For one grid, it will be either stretched or compressed based on a distortion level sampled from a uniform distribution $\mathcal{U}(-\delta, \delta)$ (line 5-8). (3) Distorted grids are then remapped to construct a new image (line 10-13). This remapping step will drop certain pixels: the compressed grids will drop rows or columns of data; the stretched grids will cause the new image to exceed the original boundary, thus the pixels being mapped outside of the original boundary will be dropped (e.g., in Fig.\ref{fig:RDG}, the grid at the lower-right corner in stage 2 is dropped in stage 3). (4) The distorted image is reshaped to the size of the original image through cropping or padding (line 14).

Due to the pixel dropping and displacement, \texttt{RDG} can bring stochastic gradients to satisfy property \#3. After the preprocessing, the transformed input is highly distinct from the original one (property \#2) but still keep the main semantics for the model to predict (property \#1).

\SetKwInput{KwParam}{Parameters}

\begin{algorithm}[]
    \caption{\emph{Random Distortion over Grids}}
    \label{algo:RDG}
    \SetNoFillComment
    \KwIn{original image $I\in \mathbb{R}^{h\times w}$}
    \KwOut{distorted image $I'\in \mathbb{R}^{h\times w}$}
    \KwParam{distortion limit $\delta\in[0,1]$; size of grid $d$.}
    \BlankLine
      
      \tcc{1.Select a starting point, e.g., upper-left corner}
      $x_0=0$, $y_0=0$\;
      \tcc{2.Random distortion over grids}
      $n_w=w//d$, $n_h=h//d$\;
      $\mathcal{G}_I=\{(x_m, y_n)|(m, n)\in\{(0, ..., n_w)\times(0, ..., n_h)\}\}$\;
      \For{$(x_m, y_n)$ in $\mathcal{G}_I\backslash\{(x_0, y_0)\}$}{
          $\delta_x\sim\mathcal{U}(-\delta, \delta)$\;
          $\delta_y\sim\mathcal{U}(-\delta, \delta)$\;
          $x_m = x_{m-1} + d\times(1+\delta_x)$\;
          $y_n = y_{n-1} + d\times(1+\delta_y)$\;
      }
      \tcc{3.Remapping grids in $I$ to $I'$}
      $\mathcal{G}_{I'}=\{(x'_m, y'_n)|x'_m=d\times m, y'_n=d\times n,(m, n)\in\{(0, ..., n_w)\times(0, ..., n_h)\}\}$\;
      \For{$(x'_m, y'_n)$ in $\mathcal{G}_{I'}\backslash\{(x'_0, y'_0)\}$}{
          $I'(x'_{m-1}:x'_m, y'_{n-1}:y'_n)=
          \texttt{Remapping}(I(x_{m-1}:x_m, y_{n-1}:y_n))$\;
      }
      \tcc{4.Reshape $I'$ to the size of $I$}
      $I'= \texttt{reshape}(I')$ s.t. $I'\in \mathbb{R}^{h\times w}$\;
      \Return $I'$\;
\end{algorithm}

\subsection{Operation $g_2(\cdot)$}

The design of this operation is relatively easier: it only needs to meet properties \# 1 and \#4. Through empirical evaluations, we identify the optimal operation:

\subsubsection{Feature Distillation (\texttt{FD})}

JPEG compression is a non-differentiable operation (property \#4) that can remove high-frequency information from the input. This operation has been demonstrated effective in increasing the model robustness and removing adversarial perturbations~\cite{yin2019fourier,das2018shield,guo2018countering} without degrading the model accuracy (property \#1). 

We select a compression-based defense, Feature Distillation (\texttt{FD}) \cite{liu2019feature}, as the operation $g_2(\cdot)$. This is an advanced JPEG compression method, which upgrades the default quantization table to better adapt to the machine's visionary behavior. \texttt{FD} is shown to be more effective towards adversarial attacks compared to other JPEG based methods \cite{liu2019feature}.

\subsection{Integration of $g_1(\cdot)$ and $g_2(\cdot)$}
\label{Case Studies}

Given the selected operations, we can integrate them to get the final solutions, which can satisfy all the properties and mitigate advanced gradient attacks. 


\begin{packeditemize}
\item \texttt{FD+Rand}: we first apply the feature distillation operation over the input. Then we send the compressed image to the randomization layer for rescaling and padding. After the transformation, the output can be fed to the DL model. 

\item \texttt{FD+RDG}: similar to the first solution, the first step is also feature distillation. Then we use \texttt{RDG} instead of \texttt{Rand} to further process the input. 

\end{packeditemize}


\noindent{\textbf{Security Analysis.}}
\label{subsec:analysis}
Both of the two end-to-end solutions use the \texttt{FD} operation, which is non-differentiable. Then the integrated functions are also non-differentiable, satisfying property \#4. The \texttt{Rand} and \texttt{RDG} operations randomize the pixels, making the solutions meet property \#3. 

We measure the model accuracy to verify the property \#1 of the two solutions. For property \#2, we measure the $l_2$ norm and Structural Similarity (SSIM) score \cite{hore2010image} between the original and transformed images. A larger $l_2$ norm or a smaller SSIM score indicates that the preprocessing function can bring a larger amount of changes to the image. Table \ref{tab: compare} reports the results and comparisons with other techniques. We can observe all those methods can maintain high accuracy for the model. \texttt{RDG}, \texttt{Rand} and their integration with \texttt{FD} can outperform other solutions with larger $l_2$ norm and smaller SSIM values. This validates that the selected operations and integrated solutions have expected properties for attack mitigation. 


\begin{table}[!htb]
\centering
\caption{Comparisons of different solutions for model accuracy and amount of changes.}
\newcommand{\tabincell}[2]{\begin{tabular}{@{}#1@{}}#2\end{tabular}}
\begin{tabular}{c c c c} 
\Xhline{1pt}
\textbf{Name of the Defense} & \textbf{$l_{2}$ norm} & \textbf{SSIM} & \textbf{Acc}\\
\Xhline{1pt}
\tabincell{c} {\texttt{FD} \cite{liu2019feature} + \texttt{RDG}} & \textbf{0.1944} & \textbf{0.3093} & 0.95\\
\hline 
\tabincell{c} {\texttt{FD} \cite{liu2019feature} + \texttt{Rand} \cite{xie2018mitigating}} & \textbf{0.2411} & \textbf{0.2820} & 0.94\\
\hline 
\tabincell{c} {\texttt{RDG}} & \textbf{0.1903} & \textbf{0.3116} & 0.96\\
\hline 
\tabincell{c} {\texttt{Rand} \cite{xie2018mitigating}} & \textbf{0.2268} & \textbf{0.3110} & 0.96\\
\hline 
\tabincell{c} {\texttt{FD} \cite{liu2019feature}} & 0.1343 & 0.4310 & 0.97\\
\hline 
\tabincell{c} {SHIELD \cite{das2018shield}} & 0.0405 & 0.8475 & 0.94\\
\hline 
\tabincell{c} {TV \cite{guo2018countering}} & 0.0338 & 0.8759 & 0.95\\
\hline 
\tabincell{c} {Bit-depth Reduction \cite{DBLP:conf/ndss/Xu0Q18}} & 0.0709 & 0.7730 & 0.92\\
\hline 
\tabincell{c} {PD \cite{prakash2018deflecting}} & 0.0147 & 0.9877 & 0.97\\
\Xhline{1pt}
\end{tabular}
\label{tab: compare}
\end{table}

\section{Evaluation}
\label{sec:eval}

\begin{figure*}[!htbp]
\centering
\subfigure[ACC per round under BPDA attack.]
{\includegraphics[width = 0.49\linewidth]{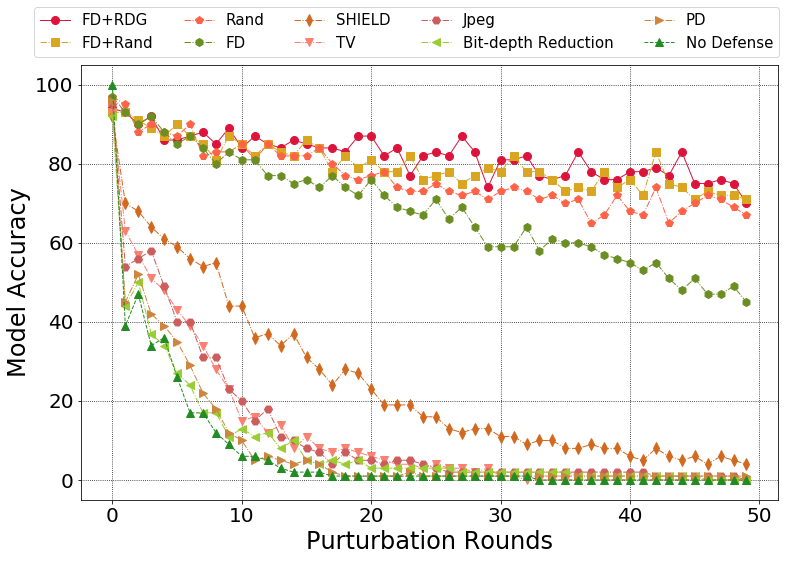}\label{BPDA_acc}}
\subfigure[ASR per round under BPDA attack.] 
{\includegraphics[width = 0.49\linewidth]{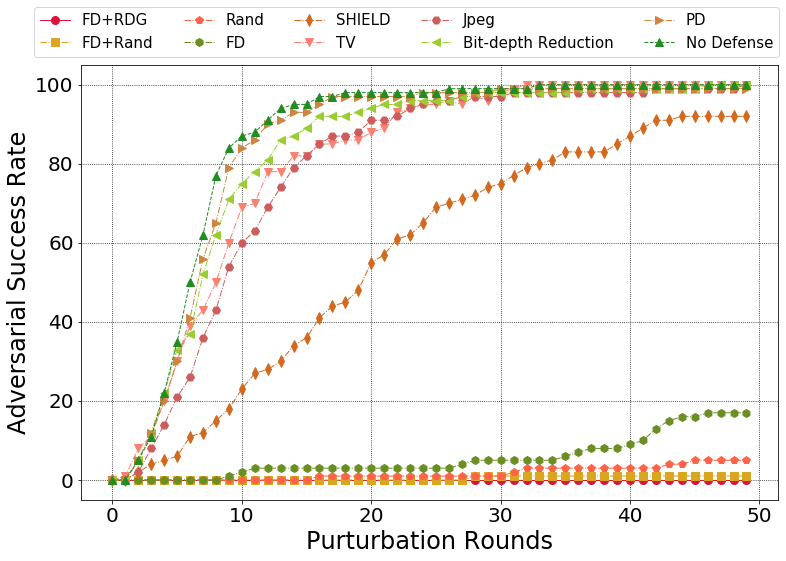}\label{BPDA_suc}}
\subfigure[ACC per round under BPDA+EOT attack.] 
{\includegraphics[width = 0.49\linewidth]{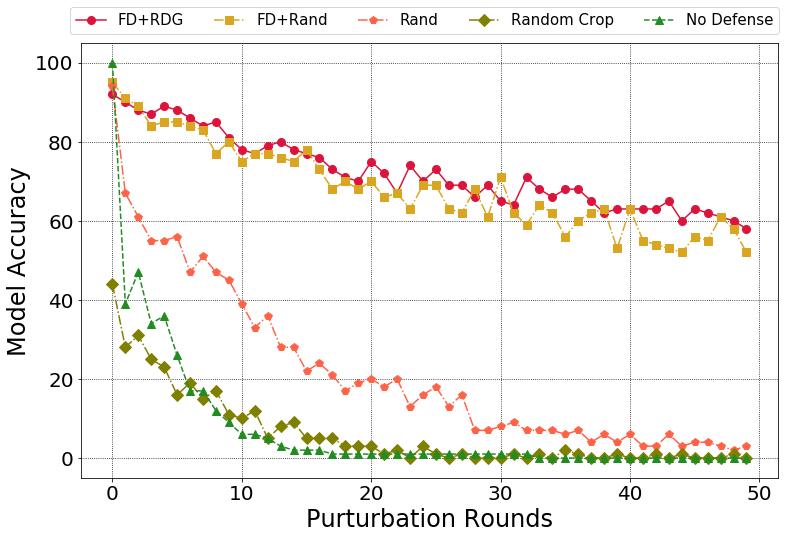}\label{EoT_acc}}
\subfigure[ASR per round under BPDA+EOT attack.] 
{\includegraphics[width = 0.49\linewidth]{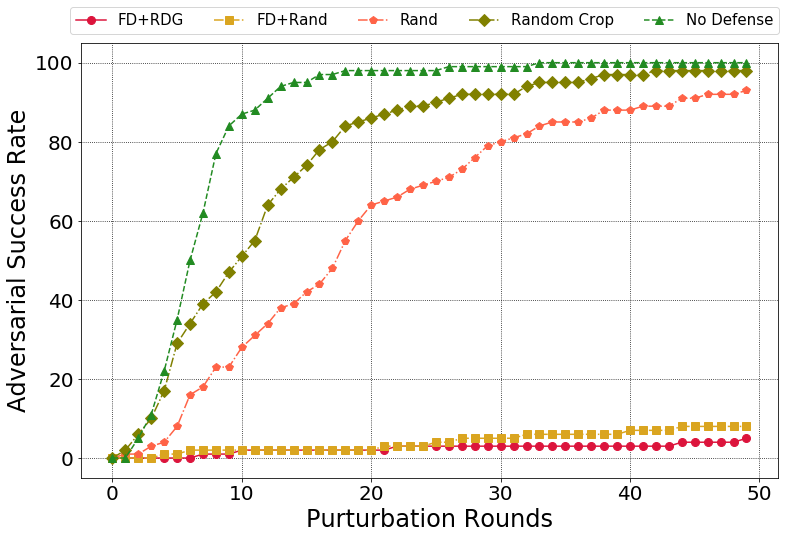}\label{EoT_suc}}
\subfigure[ACC per round under Semi-Brute-Force Attack based on EOT.] 
{\includegraphics[width = 0.49\linewidth]{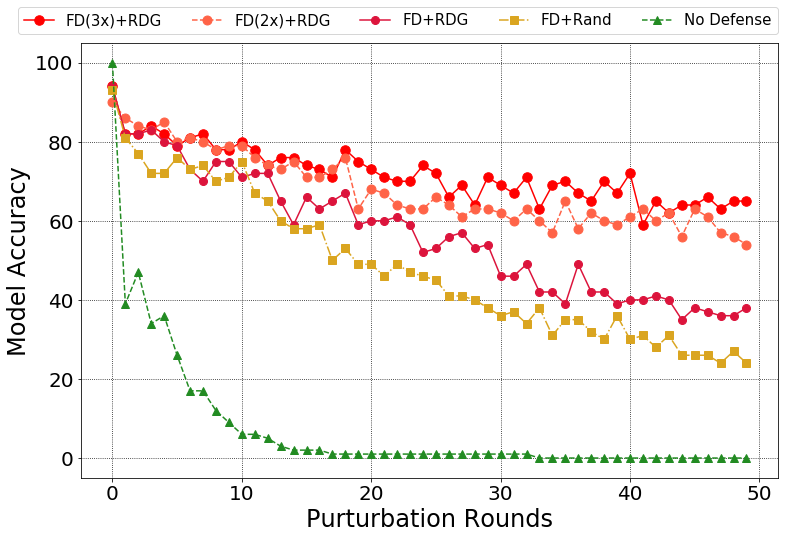}\label{newEOT_acc}}
\subfigure[ASR per round under Semi-Brute-Force Attack based on EOT.] 
{\includegraphics[width = 0.49\linewidth]{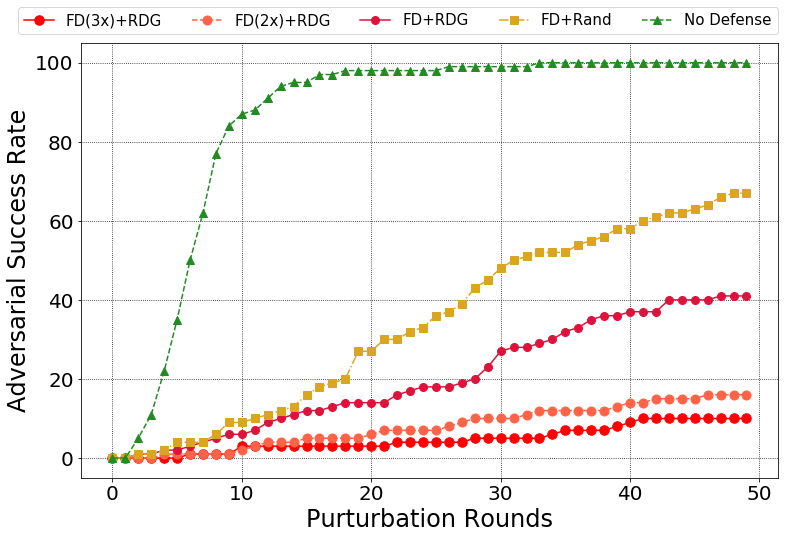}\label{newEOT_suc}}
\caption{Model accuracy and attack success rate of various techniques under different attacks. }
\label{Evaluation}
\end{figure*}

We conduct a very comprehensive evaluation of our proposed techniques. We consider various attacks: advanced gradient approximation and expectation attacks (BPDA, EOT, and their integration), as well as standard ones (I-FGSM, LBFGS, and C\&W). We compare our methods with 8 state-of-the-art solutions published in top-tier conferences over the past 2 years under the same setting. 
For the comparison, we re-implement 7 state-of-the-art solutions published in top-tier conferences recently under the same setting to show the results. Also, we give a broader comparison by including another 4 state-of-the-art solutions published in top-tier conferences recently but with more relax defense requirements. 


\subsection{Implementation}
\label{sec:setup}

\noindent{\textbf{Configurations.}}
We adopt Tensorflow as the deep learning framework to implement the attacks and defenses. The learning rate of the BPDA attack is set to 0.1, and the ensemble size of the EOT procedure is set to 30\footnote{We tested different ensemble size chosen from 2 to 40. Different ensemble size has little influence on ASR or ACC but larger ensemble size can generate AEs with less $l_{2}$.}. 
All the experiments were conducted on a server equipped with 8 Intel I7-7700k CPUs and 4 NVIDIA GeForce GTX 1080 Ti GPU.

\noindent{\textbf{Target Model and Dataset.}}
Our methods are general-purpose and can be applied to various models as a preprocessing step for computer vision tasks. Without the loss of generality, we choose a pre-trained Inception V3 model \cite{szegedy2016rethinking} over the ImageNet dataset as the target model of attacks and defenses. This state-of-the-art model can reach 78.0\% top-1 and 93.9\% top-5 accuracy. We randomly select 100 images from the ImageNet Validation dataset for AE generation. These images can be predicted correctly by this Inception V3 model. 

\noindent{\textbf{Metrics.}}
The pixel values are normalized to $[0, 1]$. We use the $l_{2}$ norm to measure the number of perturbations generated by each attack, which is calculated by computing the total root-mean-square distortion normalized by the number of pixels ($299 \times 299 \times 3$). 
We only accept adversarial examples with a $l_2$ norm smaller than 0.05. 
We consider the targeted attacks where each target label different from the correct one is randomly generated \cite{athalye2018obfuscated}. 
The BPDA and EOT attacks are iterative processes: we stop the attack when an example is generated which is predicted as its corresponding target label and the $l_2$ normal is smaller than 0.05. 
For each attack round, we measure the prediction accuracy of the generated AEs (ACC) and the attack success rate (ASR) of the targeted attack 
A higher accuracy or lower attack success rate indicates the defense is more resilient against the attacks.


\subsection{Mitigating BPDA Attack}
\label{sec:bpda}

We first evaluate the effectiveness of our two solutions against the BPDA attack. 
For comparison, we re-implemented 6 existing solutions including \texttt{FD}~\cite{liu2019feature}, SHIELD~\cite{das2018shield}, TV~\cite{guo2018countering}, JPEG~\cite{guo2018countering}, Bit-depth Reduction~\cite{DBLP:conf/ndss/Xu0Q18}, and PD~\cite{prakash2018deflecting}. 
We select these methods because they are all preprocessing-only defense, fitting our defense requirements in Section~\ref{sec:def-req}. 
Other defense solutions need to alter the target model. We give a comparison with them in Section~\ref{eva-discuss}. 
The model prediction accuracy and attack success rate in each round are shown in \figurename~\ref{BPDA_acc} and \ref{BPDA_suc}, respectively. 



We can observe that after 50 attack rounds, all the prior solutions except \texttt{FD} can only keep the model accuracy lower than 5\%, and attack success rates reach higher than 90\%. 
Those defenses fail to mitigate the BPDA attack.
\texttt{FD} can keep the attack success rate lower than 20\% and the model accuracy is around 40\%. This is better but still not very effective in maintaining the model's robustness.

In contrast, three solutions are particularly effective against BPDA attack: \texttt{FD+Rand}, \texttt{FD+RDG}, and \texttt{Rand}. 
Those techniques can maintain an acceptable model accuracy (around 70\% for 50 perturbation rounds), and restrict the attack success rate to around 0. 
These results are consistent with the $l_2$ norm and SSIM metrics in \tablename~\ref{tab: compare}: the randomization effects in those operations cause large variances between $g(x)$ and $x$, thus invalidating the BPDA attack assumption. 
Moreover, if we use the \texttt{RDG} alone as the defense method, the results of the ACC and ASR will be very similar to the one of \texttt{FD+RDG} (so using \texttt{RDG} alone as a defense is not presented in the \figurename~\ref{BPDA_acc} and \ref{BPDA_suc}). 
However, we do notice that \texttt{Rand} alone can be attacked by EOT as shown in~\cite{athalye2018obfuscated} and we found that \texttt{RDG} faces the same situation. 
More evaluation of mitigating BPDA+EOT attacks are shown in Section \ref{sec:bpda+eot}. 

We continue the attack until the samples with adversarial perturbations reach the $l_2$ bound (0.05). 
For \texttt{FD+RDG}, the adversary needs 231 rounds to reduce the model accuracy to 57\% and achieve an attack success rate of 2\%. 
For \texttt{FD+Rand}, the adversary needs 188 rounds to reduce the model accuracy to 63\% and achieve an attack success rate of 3\%. 
Therefore, we conclude that our proposed techniques can effectively mitigate the BPDA attack.

\subsection{Mitigating BPDA+EOT Attack}
\label{sec:bpda+eot}

Next, we consider a form of more powerful attack which combines BPDA and EOT~\cite{athalye2018obfuscated,tramer2020adaptive}. This attack can beat the defenses with both the shatter gradients and stochastic gradients. 
For experimentation, we only consider methods that can mitigate the BPDA attack. 
This gives us two baseline solutions: \texttt{Rand} and Random Crop\footnote{We did not report Random Crop in Section \ref{sec:bpda} as it causes a big drop to the model accuracy (30\%-40\%).} \cite{guo2018countering}. 
\figurename~\ref{EoT_acc} and \ref{EoT_suc} report model accuracy (ACC) and attack success rate (ASR) under BPDA+EOT attack.


We can observe both \texttt{Rand} and Random Crop fail to mitigate this strong attack: the model accuracy drops to below 20\% after 20 rounds, and the attack success rate reaches 100\% after 50 rounds. 
In contrast, our proposed techniques can still hold an accuracy of higher than 50\% and attack success rate of smaller than 10\% after 50 attack rounds. \texttt{FD+RDG} is slightly better than \texttt{FD+Rand}. 
The essential component that makes our solutions outperform others is the non-differentiable transformation \texttt{FD} function, which can conceal its gradients and remove the adversarial perturbation more effectively. 

We perform more rounds of attacks until the samples with adversarial perturbations reach the $l_2$ bound (0.05). 
\texttt{FD+RDG} can maintain a model accuracy of 58\% and reduce the attack success rate to 7\%, while \texttt{FD+Rand} can maintain an accuracy of 54\% and keep the attack success rate as 9\%.

\subsection{Mitigating Semi-Brute-Force Attacks based on EOT}

\label{sec:discuss}


Although the adversary cannot use EOT to attack the entire preprocessing function $g$, it is possible that he can generate AEs only based on the differentiable component. In our case, the adversary can ignore the existence of $g_2$ and perform EOT on the operation $g_1$ only. 
With the generated AE, the adversary can test if it can defeat $g_2$ as well. 
He can repeat the above procedure until a proper AE is found that can fool both $g_1$ and $g_2$. Figs~\ref{newEOT_acc} and \ref{newEOT_suc} show the ACC and ASR of this semi-brute-force attack. 
We can observe that after 50 rounds, the ACC of \texttt{FD+RDG} is 40\% and the ASR is 40\%. 
For \texttt{FD+Rand}, the ACC is 20\% and ASR is 70\%. 
After 10,000 rounds (which takes 19 hours 54 minutes), the ACC of \texttt{FD+Rand} stays at 8\% and the ASR is 89\%. 
The ACC of \texttt{FD+RDG} stays at 8\% and the ASR is 90\%.

Although this attack can break our solutions, it is not very practical. First, the cost of generating adversarial examples is too high. Second, the technique is not sophistical and optimal. EOT is used to target the randomization operation $g_1$ only. However, for the operation $g_2$ (a non-differentiable function \texttt{FD} used in this paper), the adversary has to use a brute-force fashion to handle: trying different AEs generated from EOT targeting only $g_1$ until finding AEs that can fool $g_1+g_2$. Its attack success rate is relying only on the robustness level of $g_2$: it is possible to design a stronger $g_2$ to further increase the attack cost or even totally mitigate it. For instance, in our design, if we can follow \cite{liu2019feature} to add one extra \texttt{FD} operation to make $g$ as \texttt{FD$\times 2$+RDG}, then the ACC after 50 rounds becomes around 55\% and the ASR is 17\%. 
We can further increase the robustness of $g_2$ with \texttt{FD$\times 3$} operations, as shown in Figs \ref{newEOT_acc} and \ref{newEOT_suc}, and \tablename~\ref{tab:asr_round}.

\begin{table}[!htbp]
  \centering
  \caption{Semi-brute-force attacks against different defenses.}
    \begin{tabular}{cccccc}
    \toprule
    \multirow{2}[2]{*}{Operations} & \multicolumn{5}{c}{Attak rounds required to target ASR} \\
    \cline{2-6}
          & 10\%  & 30\%  & 50\%  & 70\%  & 90\% \\
    \midrule
    
    \texttt{FD+RDG} & 14 & 38 & 59 & 207 & 7424 \\
    \texttt{FD$\times 2$+RDG} & 31 & 83 & 538 & 10000+ & 10000+ \\
    \texttt{FD$\times 3$+RDG} & 57 & 216 & 10000+ & 10000+ & 10000+ \\
    \bottomrule
    \end{tabular}
  \label{tab:asr_round}
\end{table}

\subsection{Mitigating Standard Attacks}

In addition to the advanced attacks, we also test our defenses against standard attacks, i.e., I-FGSM, LBFGS, and C\&W. 
For I-FGSM, AEs are generated under $l_{\infty}$ constraint of 0.03. 
For LBFGS and C\&W, the attack process is iterated under $l_2$ constraint and stops when all targeted AEs are found. 
We measure the model accuracy (ACC) and attack success rate (ASR) with the protection of \texttt{FD+Rand} and \texttt{FD+RDG}. 

The results are shown in~\tablename~\ref{table:attacks}. 
For benign samples only, our proposed techniques have little influence on the model accuracy (92\% and 95\% for \texttt{FD+Rand} and \texttt{FD+RDG}, respectively). 
For defeating AEs generated by these standard attacks, the attack success rate can be kept as 0\% and the model accuracy can be maintained as around 90\%. 
We provide more experimental details with different attack settings and defense configurations in Appendix~\ref{appendix}. 

\begin{table}[!htbp]
\centering
\caption{Performance of our defenses against I-FGSM, LBFGS, and C\&W.}
\begin{tabular}{cccccccc}
\hline
\multirow{2}{*}{Attack} & \multirow{2}{*}{$l_2$} & \multicolumn{2}{c}{\textbf{No Defense}} & \multicolumn{2}{c}{\texttt{FD+Rand}} & \multicolumn{2}{c}{\texttt{FD+RDG}} \\
& & ACC & ASR & ACC & ASR & ACC & ASR \\ \hline
No attack & 0.0 & 100\% & Nan & 92\% & Nan & 95\% & Nan \\
I-FGSM & 0.010 & 2\% & 95\% & 88\% & 0\% & 93\% & 0\% \\
LBFGS & 0.001 & 0\% & 100\% & 92\% & 0\% & 91\% & 0\% \\
C\&W & 0.016 & 0\% & 100\% & 86\% & 0\% & 87\% & 0\% \\ \hline
\end{tabular}
\label{table:attacks}
\end{table}

\subsection{A Broader Comparison with More Defenses}\label{eva-discuss}

\begin{table*}[!htbp]
  \centering
  \caption{Comparisons with a broader set of defenses against bounded adaptive attacks. (* denotes adversarial training).}
    \begin{tabular}{cccccccc}
    \toprule
    Defense & Attack & \#Property 1 & \#Property 2 & \#Property 3 & \#Property 4 & $l_\infty$ bounded (0.031) & $l_2$ bounded (0.05) \\
    \midrule
    Rand~\cite{xie2018mitigating}  & BPDA+EOT &  $\bullet$     &  $\bullet$     & $\bullet$      &       & 0\%   & - \\
    PixelDefend~\cite{song2017pixeldefend} & BPDA  & $\bullet$      &       &       & $\bullet$      & 9\%   &  -\\
    Crop*~\cite{guo2018countering}  & BPDA+EOT   &       & $\bullet$      & $\bullet$      &       & -     & 0\% \\
    JPEG*~\cite{guo2018countering}  & BPDA  & $\bullet$      &       &       & $\bullet$      & -     & 0\% \\
    TV*~\cite{guo2018countering}    & BPDA+EOT & $\bullet$      &       & $\bullet$      & $\bullet$      & -     & 0\% \\
    Quilting*~\cite{guo2018countering} & BPDA+EOT & $\bullet$      &       & $\bullet$      & $\bullet$      & -     & 0\% \\
    SHIELD*~\cite{das2018shield} & BPDA & $\bullet$      &       &      & $\bullet$      & -     & 0\% \\
    Pixel Deflection~\cite{prakash2018deflecting} & BPDA  & $\bullet$      &       & $\bullet$      & $\bullet$      & 0\%   & - \\
    Guided Denoiser~\cite{liao2018defense} & BPDA  & $\bullet$      &       &       & $\bullet$      & -     & 0\% \\
    ME-Net*~\cite{yang2019me} & BPDA+EOT &      & $\bullet$      & $\bullet$      & $\bullet$      & 13\% & - \\
    FD~\cite{liu2019feature} & BPDA & $\bullet$      & $\bullet$  &      & $\bullet$      & - & 10\% \\
    FD+RDG (Ours) & BPDA+EOT & $\bullet$      & $\bullet$      & $\bullet$      & $\bullet$      & -     & 58\% \\
    \bottomrule
    \end{tabular}%
  \label{tab:bounded_attack}%
\end{table*}%

In addition to the eight existing preprocessing solutions, we also compare our defenses with a broader set of defenses against bounded adaptive attacks. These methods are also based on preprocessing, and some of them are also combined with model retraining (ME-Net) or adversarial training (Crop, JPEG, TV, Quilting, ME-Net). These methods were proven to be broken partially or entirely by advanced attacks (BPDA or BPDA+EOT).

We summarize the analytic results, experimental data as well as conclusions from literature, as shown in \tablename~\ref{tab:bounded_attack}. The perturbation criteria is chosen either under $l_\infty$ bound (0.031) or $l_2$ bound (0.05) to generate imperceptible AEs. We list the model accuracy when the samples with adversarial perturbations are under the bound. We can observe that our defense shows more resistance against advanced attacks and keeps the model accuracy as high as 58\% under the $l_2$ bound.

We use the four properties (Section~\ref{sec:overview}) to reveal the reason that our solution can beat others. As shown in \tablename~\ref{tab:bounded_attack}, all the prior defenses satisfy only part of the properties. In this case, even combined with adversarial training, most defenses cannot provide enough robustness. Note that ME-Net has the best performance among these past defenses, as it meets properties \#2 \#3 \#4. However, it uses model retraining to achieve higher accuracy on clean samples, which violates the property \#1. This results in changes in model parameters, which could potentially provide more information to the adversary under the white-box setting.

From these results, we conclude that the four properties are indeed helpful to increase the difficulty of adversarial attacks. We also think there might probably be other potential properties that are useful to further improve the defenses. We encourage researchers to consider these properties when designing a preprocessing-based defense and also to find more useful properties.



\section{Future Work}





We hope that our findings in this paper can inspire both the adversary and defender to understand the adversarial examples and enhance their techniques to continue the arms races. 
We identify some possible future research directions. 

\noindent{\textbf{Defense.}} As we demonstrated in Section \ref{sec:discuss}, an semi-brute-force attack can break \texttt{FD+RDG}, although the attack cost is high. To thwart this attack, we add more \texttt{FD} operations into the preprocessing function, as suggested by \cite{liu2019feature}. In the future, we aim to design more lightweight non-differentiable operations that are infeasible to be fooled by AEs from the EOT attack. 

Our preprocessing function consists of several operations in order to satisfy all the properties in Section \ref{sec:overview}. So an adaptive attack can ignore certain operations, and only target the rest ones to generate AEs. 
In the future, we aim to design a single preprocessing function with the four properties. 
So the attacker cannot break it into different parts and attack it adaptively. 



\noindent{\textbf{Attack.}}
It is also interesting to enhance the attacks from the adversarial perspective. 
The first direction is to optimize the EOT attack and reduce the cost. 
Currently, it simply generates AEs from the randomization operation and tests if it can fool the non-differentiable operation. 
It is necessary to consider the non-differentiable operation as well and generate AEs targeting it with fewer rounds. The Brute-force fashion EOT attack is not a promising research direction. 

The second direction is to consider the preprocessing function with more than two operations or a single operation, as described above. For three or more operations, how can the adversary classify them into attacking group and ignored group needs to be explored. For one operation, a new attack strategy to deal with all four properties is necessary.


\section{Conclusion}

In this paper, we propose new solutions to mitigate the advanced gradient-based adversarial attacks (BPDA and EOT). Specifically, we first identify four properties to reveal the possible defense opportunities. Following these properties, we design two preprocessing solutions to invalidate these attacks. We evaluate our methodologies over existing attack techniques and compare with 11 state-of-the-art defense approaches. Empirical results indicate that these two proposed methods have the best performance in mitigating the powerful attacks. 

We expect the properties and solutions proposed in this paper can inspire more advanced attack and defense techniques in the future. All the efforts can enhance people's understanding about the attack and defense mechanisms, as well as the robustness of DL models. 



\bibliographystyle{IEEEtran}
\bibliography{ref} 

\appendix
\label{appendix}

\begin{table*}[h]
  \centering
  \caption{Impact of padding size on defense performance of \textbf{FD+Rand}}
    \begin{tabular}{ccccccccccc}
    \toprule
    \multirow{2}[1]{*}{Attack} & \multicolumn{2}{c}{$p=300$} & \multicolumn{2}{c}{$p=350$} & \multicolumn{2}{c}{$p=400$} & \multicolumn{2}{c}{$p=450$} & \multicolumn{2}{c}{$p=500$} \\
          & ACC   & ASR   & ACC   & ASR   & ACC   & ASR   & ACC   & ASR   & ACC   & ASR \\
    Clean & 0.95  & Nan   & 0.92  & Nan   & \textbf{0.92} & Nan   & 0.91  & Nan   & 0.89  & Nan \\
    FGSM ($\epsilon=0.01$) & 0.62  & 0.00  & 0.71  & 0.00  & \textbf{0.79} & 0.00  & 0.75  & 0.00  & 0.72  & 0.00 \\
    FGSM ($\epsilon=0.03$) & 0.50  & 0.00  & 0.54  & 0.00  & 0.58  & 0.00  & \textbf{0.62} & 0.00  & 0.60  & 0.00 \\
    I-FGSM ($\epsilon=0.01$) & 0.90  & 0.00  & \textbf{0.92} & 0.00  & 0.91  & 0.00  & 0.86  & 0.00  & 0.87  & 0.00 \\
    I-FGSM ($\epsilon=0.03$) & 0.85  & 0.00  & 0.88  & 0.00  & \textbf{0.88} & 0.00  & 0.84  & 0.00  & 0.86  & 0.00 \\
    LBFGS & \textbf{0.94} & 0.00  & 0.93  & 0.00  & 0.92  & 0.00  & 0.93  & 0.00  & 0.85  & 0.00 \\
    CW    & 0.88  & 0.00  & 0.85  & 0.00  & \textbf{0.86} & 0.00  & 0.91  & 0.00  & 0.88  & 0.00 \\
    \bottomrule
    \end{tabular}
  \label{tab:padding_size}
\end{table*}

\begin{table*}[htbp]
  \centering
  \caption{Impact of distortion limits on defense performance of \textbf{FD+RDG}}
    \begin{tabular}{ccccccccccccccc}
    \toprule
    \multirow{2}[2]{*}{Attack} & \multicolumn{2}{c}{$\delta=0.01$} & \multicolumn{2}{c}{$\delta=0.05$} & \multicolumn{2}{c}{$\delta=0.10$} & \multicolumn{2}{c}{$\delta=0.15$} & \multicolumn{2}{c}{$\delta=0.20$} & \multicolumn{2}{c}{$\delta=0.25$} & \multicolumn{2}{c}{$\delta=0.30$} \\
          & ACC   & ASR   & ACC   & ASR   & ACC   & ASR   & ACC   & ASR   & ACC   & ASR   & ACC   & ASR   & ACC   & ASR \\
    \midrule
    Clean & 0.95  & Nan   & 0.96  & Nan   & 0.95  & Nan   & 0.95  & Nan   & \textbf{0.96} & Nan   & 0.93  & Nan   & 0.91  & Nan \\
    FGSM ($\epsilon=0.01$)  & 0.70  & 0.00  & 0.66  & 0.00  & 0.69  & 0.00  & 0.73  & 0.00  & 0.69  & 0.00  & \textbf{0.75} & 0.00  & 0.72  & 0.00 \\
    FGSM ($\epsilon=0.03$)   & 0.51  & 0.00  & 0.51  & 0.00  & 0.51  & 0.00  & 0.53  & 0.00  & 0.55  & 0.00  & 0.55  & 0.00  & \textbf{0.62} & 0.00 \\
    I-FGSM ($\epsilon=0.01$)  & 0.96  & 0.00  & 0.05  & 0.00  & 0.93  & 0.00  & 0.89  & 0.00  & 0.90  & 0.00  & 0.91  & 0.00  & \textbf{0.93} & 0.00 \\
    I-FGSM ($\epsilon=0.03$) & 0.88  & 0.01  & 0.90  & 0.00  & 0.86  & 0.00  & \textbf{0.93} & 0.00  & 0.92  & 0.00  & 0.89  & 0.00  & 0.89  & 0.00 \\
    LBFGS & 0.95  & 0.00  & \textbf{0.97} & 0.00  & 0.93  & 0.00  & 0.91  & 0.00  & 0.94  & 0.00  & 0.94  & 0.00  & 0.88  & 0.00 \\
    C\&W     & 0.86  & 0.00  & 0.87  & 0.00  & 0.85  & 0.00  & \textbf{0.87} & 0.00  & 0.83  & 0.00  & 0.83  & 0.00  & 0.84  & 0.00 \\
    \bottomrule
    \end{tabular}%
  \label{tab:distort_limit}%
\end{table*}%

\begin{table*}[htbp]
  \centering
  \caption{Performance of different defenses against standard attacks}
    \begin{tabular}{ccccccccccccc}
    \toprule
    \multirow{2}[2]{*}{Attack} & \multicolumn{2}{c}{\textbf{Baseline}} & \multicolumn{2}{c}{\textbf{FD}} & \multicolumn{2}{c}{\textbf{RDG}} & \multicolumn{2}{c}{\textbf{Rand}} & \multicolumn{2}{c}{\textbf{FD+RDG}} & \multicolumn{2}{c}{\textbf{FD+Rand}} \\
          & ACC   & ASR   & ACC   & ASR   & ACC   & ASR   & ACC   & ASR   & ACC   & ASR   & ACC   & ASR \\
    \midrule
    Clean & 1.00  & Nan   & \textbf{0.97} & Nan   & 0.96  & Nan   & 0.96  & Nan   & 0.95  & Nan   & 0.92  & Nan \\
    FGSM ($\epsilon=0.01$)  & 0.36  & 0.00  & 0.64  & 0.00  & 0.56  & 0.00  & 0.60  & 0.00  & 0.73  & 0.00  & \textbf{0.79} & 0.00 \\
    FGSM ($\epsilon=0.03$) & 0.39  & 0.00  & 0.47  & 0.00  & \textbf{0.66} & 0.00  & 0.60  & 0.00  & 0.53  & 0.00  & 0.58  & 0.00 \\
    I-FGSM ($\epsilon=0.01$) & 0.13  & 0.79  & \textbf{0.96} & 0.00  & 0.92  & 0.00  & 0.92  & 0.00  & 0.89  & 0.00  & 0.91  & 0.00 \\
    I-FGSM ($\epsilon=0.03$) & 0.02  & 0.95  & 0.87  & 0.00  & 0.75  & 0.00  & 0.86  & 0.01  & \textbf{0.93} & 0.00  & 0.88  & 0.00 \\
    LBFGS & 0.00  & 1.00  & \textbf{0.97} & 0.00  & 0.92  & 0.00  & 0.95  & 0.00  & 0.91  & 0.00  & 0.92  & 0.00 \\
    CW    & 0.00  & 1.00  & 0.84  & 0.00  & 0.83  & 0.00  & 0.86  & 0.00  & \textbf{0.87} & 0.00  & 0.86  & 0.00 \\
    \bottomrule
    \end{tabular}%
  \label{tab:attack_append}%
\end{table*}%

We evaluate the robustness of our defenses against standard attacks, FGSM, I-FGSM, LBFGS, C\&W. All attacks are conducted as targeted attack, we randomly select classes that are different from original ones. An attack succeeds only if the prediction of model becomes the targeted class. We use Cleverhans package \cite{papernot2018cleverhans} to generate AE of all standard attacks. For FGSM and I-FGSM, AEs are generated under two different $l_\infty$ constraints ($\epsilon=0.01, 0.03$). I-FGSM is iteratied 10 times. For LBFGS and C\&W, the optimization process is iterated until all targeted AEs are found under $l_2$ constraint. For LBFGS, the binary search steps are set to 5 and the maximum number of iterations is set to 1000. For CW, the binary search steps are set to 5, the maximum number of iterations is set to 1000, and the learning rate is 0.1. We evaluate the model accuracy (ACC) and attack success rate (ASR), as well as the $l_\infty$ norm and $l_2$ norm, \tablename~\ref{tab:attacks_baseline}. Note that FGSM is just a one-step attack and it is not really effective as a targeted attack. Its iterative version I-FGSM with $\epsilon=0.03$ can reach ASR 95\%. Two optimization-based attacks, LBFGS and CW, can even entirely break the baseline model with 100\% ASR.

\begin{table}[htbp]
  \centering
  \caption{Standard attacks on baseline model. }
    \begin{tabular}{ccccc}
	\toprule
    \multirow{2}[1]{*}{Attack} & \multirow{2}[1]{*}{$l_{\infty}$} & \multirow{2}[1]{*}{$l_2$} & \multicolumn{2}{c}{\textbf{Baseline}} \\
          &       &       & ACC   & ASR \\
    \midrule
    Clean & 0.000 & 0.0000 & 1.00  & Nan \\
    FGSM ($\epsilon=0.01$) & 0.010 & 0.0099 & 0.36  & 0.00 \\
    FGSM ($\epsilon=0.03$) & 0.030 & 0.0294 & 0.39  & 0.00 \\
    I-FGSM ($\epsilon=0.01$) & 0.010 & 0.0040 & 0.13  & 0.79 \\
    I-FGSM ($\epsilon=0.03$) & 0.030 & 0.0098 & 0.02  & 0.95 \\
    LBFGS & 0.021 & 0.0013 & 0.00  & 1.00 \\
    CW    & 0.156 & 0.0162 & 0.00  & 1.00 \\
    \bottomrule
    \end{tabular}%
  \label{tab:attacks_baseline}%
\end{table}%

For the operations proposed in Section \ref{sec:technique}, there are several hyper-parameters to be determined. 

For \texttt{FD+Rand}, the padding size $p$ and padding value $p_{value}$ have influence on the image randomization level. We first vary the padding size from 300 to 500, as shown in \tablename~\ref{tab:padding_size}. The optimal padding size is 400. Then we test the padding value for 0, 0.5 and 1, as shown in \tablename~\ref{tab:padding_value}.

\begin{table}[htbp]
  \centering
  \caption{Impact of padding value on defense performance of \textbf{FD+Rand}}
    \begin{tabular}{ccccccc}
    \toprule
    \multirow{2}[2]{*}{Attack} & \multicolumn{2}{c}{$p_{value}=0$} & \multicolumn{2}{c}{$p_{value}=0.5$} & \multicolumn{2}{c}{$p_{value}=1$} \\
          & ACC   & ASR   & ACC   & ASR   & ACC   & ASR \\
    \midrule
    Clean & 0.91  & Nan   & 0.92  & Nan   & \textbf{0.93} & Nan \\
    FGSM ($\epsilon=0.01$) & 0.73  & 0.00  & \textbf{0.79} & 0.00  & 0.75  & 0.00 \\
    FGSM ($\epsilon=0.03$) & 0.55  & 0.00  & 0.58  & 0.00  & \textbf{0.60} & 0.00 \\
    I-FGSM ($\epsilon=0.01$) & 0.89  & 0.00  & 0.91  & 0.00  & \textbf{0.91} & 0.00 \\
    I-FGSM ($\epsilon=0.03$) & 0.87  & 0.00  & \textbf{0.88} & 0.00  & 0.83  & 0.00 \\
    LBFGS & 0.91  & 0.00  & \textbf{0.92} & 0.00  & 0.87  & 0.00 \\
    CW    & 0.88  & 0.00  & 0.86  & 0.00  & \textbf{0.91} & 0.00 \\
    \bottomrule
    \end{tabular}%
  \label{tab:padding_value}%
\end{table}%

For \texttt{FD+RDG}, the distortion limit $\delta$ has influence on distortion level of each grids. It also affects the ratio of pixels that will be dropped. We apply a linear search of $\delta$ from 0.01 to 0.30, as show in \tablename~\ref{tab:distort_limit}. The ASR becomes 0\% under our defenses, which shows that the adversarial perturbation is delicate to this kind of distortion. A larger $\delta$ decreases the ACC on clean examples. Thus, a moderate $\delta=0.15$ is chosen as the optimal value. 

Finally, with chosen hyper-parameters, we test the performance of different defenses against standard attacks, \texttt{FD}, \texttt{RDG}, \texttt{FD+RDG}, \texttt{FD+Rand}, in \tablename~\ref{tab:attack_append}.

\end{document}